\documentclass[10pt,conference]{IEEEtran}

\usepackage{extarrows}

\usepackage{tabularx}
\usepackage{booktabs}
\usepackage{multicol}
\usepackage{multirow}
\usepackage{threeparttable}
\usepackage{amsmath,amssymb,amsfonts}
\usepackage{algorithmic}
\usepackage{graphicx}
\usepackage{textcomp}
\usepackage{xcolor}
\def\BibTeX{{\rm B\kern-.05em{\sc i\kern-.025em b}\kern-.08em
		T\kern-.1667em\lower.7ex\hbox{E}\kern-.125emX}}

\usepackage{caption}
\usepackage{subcaption}
\usepackage{balance} 
\usepackage{xspace}
\usepackage[T1]{fontenc}
\usepackage[scaled=0.81]{beramono}
\usepackage{balance}

\usepackage{enumitem}  
\usepackage{listings}
\usepackage{url}
\usepackage[ruled,lined,linesnumbered,vlined]{algorithm2e}
\usepackage{multirow}
\usepackage{color}

\usepackage{listings, xcolor}

\definecolor{verylightgray}{rgb}{.97,.97,.97}

\lstdefinelanguage{Solidity}{
	keywords=[1]{anonymous, assembly, assert, balance, break, call, callcode, case, catch, class, constant, continue, contract, debugger, default, delegatecall, delete, do, else, emit, event, export, external, false, finally, for, function, gas, if, implements, import, in, indexed, instanceof, interface, internal, is, length, library, log0, log1, log2, log3, log4, memory, modifier, new, payable, pragma, private, protected, public, pure, push, require, return, returns, revert, selfdestruct, send, storage, struct, suicide, super, switch, then, this, throw, transfer, true, try, typeof, using, value, view, while, with, addmod, ecrecover, keccak256, mulmod, ripemd160, sha256, sha3}, 
	keywordstyle=[1]\color{blue}\bfseries,
	keywords=[2]{address, bool, byte, bytes, bytes1, bytes2, bytes3, bytes4, bytes5, bytes6, bytes7, bytes8, bytes9, bytes10, bytes11, bytes12, bytes13, bytes14, bytes15, bytes16, bytes17, bytes18, bytes19, bytes20, bytes21, bytes22, bytes23, bytes24, bytes25, bytes26, bytes27, bytes28, bytes29, bytes30, bytes31, bytes32, enum, int, int8, int16, int24, int32, int40, int48, int56, int64, int72, int80, int88, int96, int104, int112, int120, int128, int136, int144, int152, int160, int168, int176, int184, int192, int200, int208, int216, int224, int232, int240, int248, int256, mapping, string, uint, uint8, uint16, uint24, uint32, uint40, uint48, uint56, uint64, uint72, uint80, uint88, uint96, uint104, uint112, uint120, uint128, uint136, uint144, uint152, uint160, uint168, uint176, uint184, uint192, uint200, uint208, uint216, uint224, uint232, uint240, uint248, uint256, var, void, ether, finney, szabo, wei, days, hours, minutes, seconds, weeks, years},	
	keywordstyle=[2]\color{teal}\bfseries,
	keywords=[3]{block, blockhash, coinbase, difficulty, gaslimit, number, timestamp, msg, data, gas, sender, sig, value, now, tx, gasprice, origin},	
	keywordstyle=[3]\color{violet}\bfseries,
	identifierstyle=\color{black},
	sensitive=false,
	comment=[l]{//},
	morecomment=[s]{/*}{*/},
	commentstyle=\color{gray}\ttfamily,
	stringstyle=\color{red}\ttfamily,
	morestring=[b]',
	morestring=[b]"
}

\definecolor{darkgreen}{rgb}{0.0, 0.5, 0.13}

\newcommand{\tool}{\hbox{\textsc{BlockEye}}\xspace}
\newcommand{\ethereum}{\hbox{Ethereum}\xspace}
\newcommand{\defi}{\hbox{DeFi}\xspace}

\newcommand{\myparagraph}[1]{\vspace*{0.14cm}\noindent\textbf{\emph{#1.}}\quad}

\newcommand{\etal}{\hbox{\emph{et al.}}\xspace}
\newcommand{\eg}{\hbox{\emph{e.g.}}\xspace}
\newcommand{\ie}{\hbox{\emph{i.e.}}\xspace}

\newcommand{\etc}{\hbox{\emph{etc.}}\xspace}

\newcommand{\videoUrl}{\texttt{\url{https://youtu.be/7DjsWBLdlQU}}}

\AtBeginDocument{%
  \providecommand\BibTeX{{%
    \normalfont B\kern-0.5em{\scshape i\kern-0.25em b}\kern-0.8em\TeX}}}  

\title{\textsc{BlockEye}: Hunting For DeFi Attacks\\on Blockchain}

\author{\IEEEauthorblockN{
Bin Wang, Han Liu, Chao Liu, Zhiqiang Yang, Qian Ren, Huixuan Zheng, Hong Lei}
\IEEEauthorblockA{
\textit{Oxford-Hainan Blockchain Research Institute}, Hainan, China\\}
}

\begin{document}
	\maketitle

\begin{abstract}

Decentralized finance, \ie, DeFi, has become the most popular type of 
application on many public blockchains (\eg, \ethereum) in recent years. 
Compared to the traditional finance, \defi allows customers 
to flexibly participate in diverse blockchain financial services (\eg, lending, borrowing, 
collateralizing, exchanging \etc) via smart contracts at a relatively low cost of trust. 
However, the open nature of \defi inevitably introduces a large attack surface, 
which is a severe threat to the security of participants' funds. In this paper, 
we proposed \tool, a real-time attack detection system for \defi projects on 
the \ethereum blockchain. Key capabilities provided by \tool are twofold: (1) 
Potentially vulnerable \defi projects are identified based on an automatic 
security analysis process, which performs symbolic reasoning on the data flow of 
important service states, \eg, asset price, and checks whether they can be 
externally manipulated. (2) Then, a transaction monitor is installed off-chain for 
a vulnerable \defi project. Transactions sent not only to that project but other 
associated projects as well are collected for further security analysis. 
A potential attack is flagged if a violation is detected on a critical invariant 
configured in \tool, \eg, Benefit is achieved within a very short time and way much 
bigger than the cost. We applied \tool in several popular \defi projects and managed 
to discover potential security attacks that are unreported before. A video of \tool 
is available at \videoUrl.

\end{abstract}

\begin{IEEEkeywords}
	DeFi, oracle analysis, attack monitoring
\end{IEEEkeywords}




	
	\section{Introduction}
	\label{sec:intro}
	Recent years have witnessed a rapid growth of decentralized finance application, or \defi application, on the public blockchain ecosystem, \eg, \ethereum~\cite{Wood2014Ethereum}.
Unlike in traditional finance, \defi applications leverage the transparency and openness nature of decentralized network (\ie, blockchain) to provide a diversity range of financial services, \eg, lending, borrowing, collateralizing, exchanging \etc, all without trust or dependency on third-party intermediaries.

\begin{figure}[h]
\centering
\includegraphics[width=\linewidth]{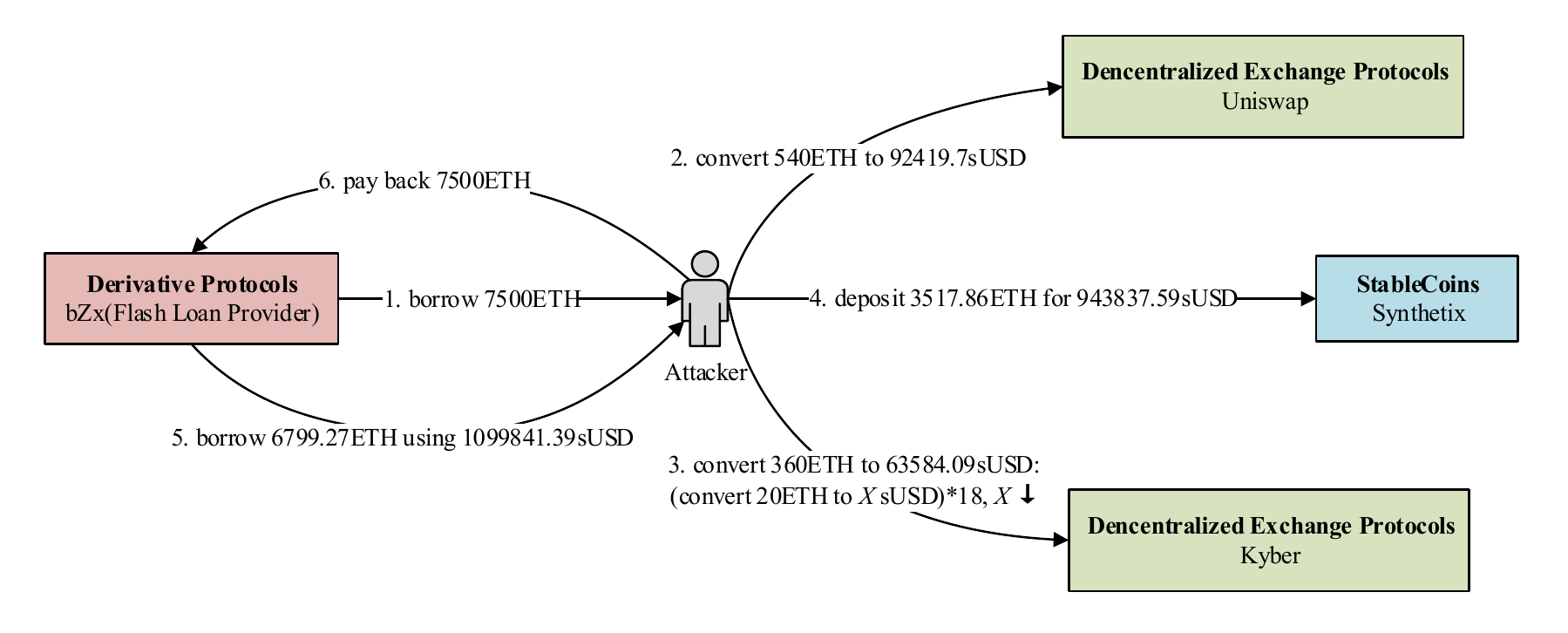}
\caption{\label{fig:bzx-attack}Attack on the \texttt{bZx} project.}
\end{figure}

While \defi has been gaining an increasing level of market growth in terms of both popularity and liquidity, its openness nature also leaves a large space to external attacks, which may severely threaten the security of \defi participants' funds.
To elaborate on this point, consider a real-world attack (see Figure~\ref{fig:bzx-attack}) on \texttt{bZx} project, which is a \defi project for lending and margin trading on \ethereum.
In this case, the attacker leveraged an oracle dependency of \texttt{bZx} on other \defi projects (\ie, \texttt{Uniswap} and \texttt{Kyber}) in manipulating cryptoasset exchange rates, making net profit with a single atomic transaction.

Specifically, as shown in Figure~\ref{fig:bzx-attack}, the attacker launched a sequence of six internal transactions, consisting of borrowing (\eg, transaction 1 and 5), exchanging (\eg, transaction 2, 3, and 4), and paying back (\eg, transaction 6) cryptoassets (\ie, \texttt{ETH} and \texttt{sUSD}).
Note, these transactions are packed into a single external transaction in the exact order as in Figure~\ref{fig:bzx-attack}, which is then executed atomically by \ethereum.
In its execution, the attacker first borrowed $7,500$ \texttt{ETH} from \texttt{bZx} (transaction 1), then used $4,417.86$ borrowed \texttt{ETH} in exchange of \texttt{sUSD} with other \defi projects (\ie, \texttt{Uniswap}, \texttt{Kyber}, and \texttt{Synthetix} in transactions 2--4).
Because \texttt{bZx} relies on \texttt{Uniswap} and \texttt{Kyber} for price feed oracles, which are instead susceptible to large amount transactions, the attacker can therefore largely skewed exchange rate of \texttt{ETH/sUSD} in \texttt{bZx} in favour of him or herself.
After that, he or she triggered transaction 5, borrowing $6,799.27$ \texttt{ETH} with all holding \texttt{sUSD} (\ie, $1,099,841.39$), followed by a last transaction 6 in paying back $7,500$ borrowed \texttt{ETH} at the very beginning.
The outcome of transactions 1--6 is thus a net profit of $2,381.41$ \texttt{ETH} (minus a small amount of \texttt{ETH} for paying gas fee\cite{Wood2014Ethereum}), or $\text{\$}600\text{K}$, for the attacker.

We point out the crux of this kind of arbitrage (\ie, making profits by buying and selling goods at different prices) is for the attacker successfully controlling exchange rates of cryptoasset pairs, \texttt{ETH/sUSD} here in Figure~\ref{fig:bzx-attack}, by exploiting data dependencies of \texttt{bZx} on \texttt{Uniswap} and \texttt{Kyber}.


\begin{figure*}
	\centering
	\includegraphics[width=.73\linewidth]{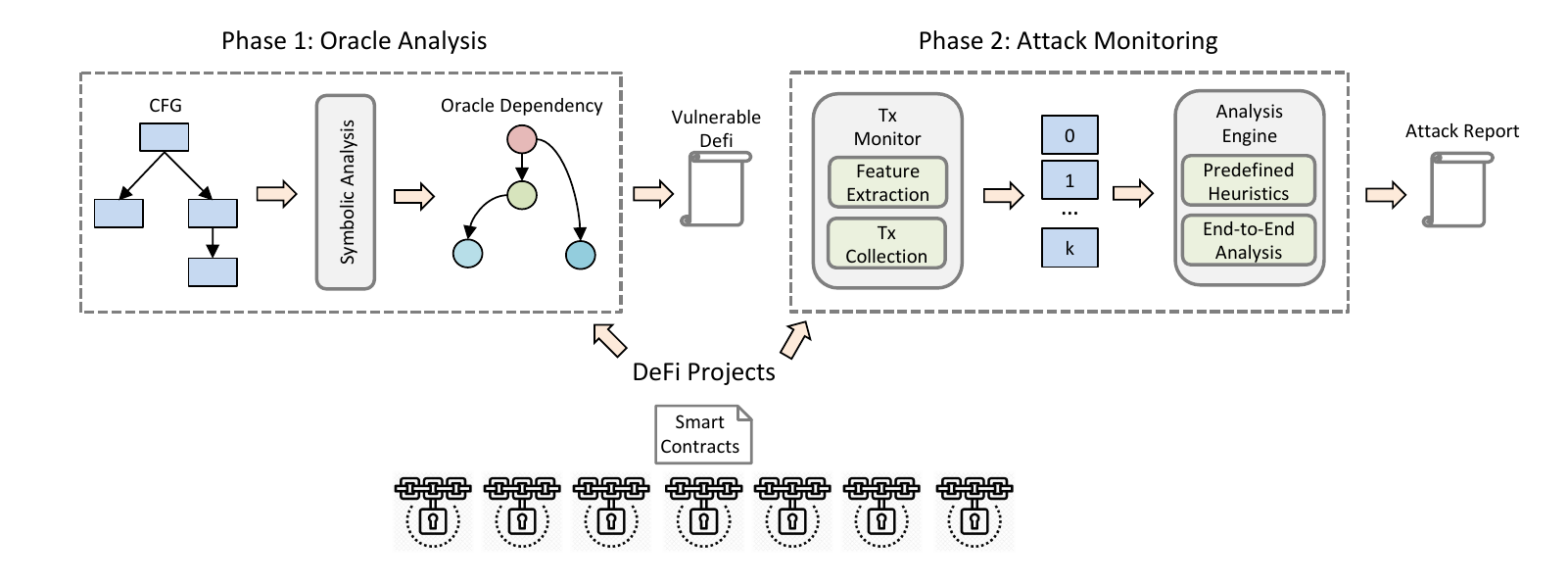}
	\caption{\label{fig:framework}%
		The general workflow of \tool.}
\end{figure*}

While many previous research works and tools have focused on the security of 
smart contracts~\cite{Luu2016Making,tsankov2018securify,liu2018reguard,liu2018s}, there is 
relatively little study on the security of \defi projects as aforementioned. In general, 
detection of these attacks requires a deep understanding on both the business nature of a \defi project
as well as the market it is involved in, which are missing in existing solutions to finding 
low-level bugs of smart contracts. We summarized challenges in terms of addressing 
security problems in \defi projects as below.

\myparagraph{Challenge 1: Model DeFi Dependency}
Attacks on \defi often involve multiple projects rather than a single one. 
Therefore, detection of such attacks requires an effective modeling of critical 
dependencies among \defi projects, \eg, information flow between two \defi. 
While a full analysis would introduce too much complexity, an abstract analysis 
might not be able to capture important high-level business semantics.

\myparagraph{Challenge 2: Understand End-To-End Transactions}
Furthermore, whether a sequence of transactions is considered as malicious 
is largely determined by end-to-end analysis, \ie, comparing benefits and costs 
in the transaction sequence. However, such insights are hard to configure and generate 
based on existing analysis infrastructures for blockchains.


\myparagraph{The \tool Solution}
To overcome above challenges, we have designed and developed \tool, the very first 
automatic attack detection platform for blockchain \defi projects. The key insights 
behind \tool are twofold. First, a symbolic analysis is performed in \tool 
to reason on important data flow (\eg, asset price) among associated \defi projects. 
Potentially vulnerable projects are identified in this process. Then, \tool installs 
a runtime monitor on vulnerable \defi projects to detect potential attacks on the fly. Specifically, 
an end-to-end economic analysis is executed to report malicious transactions based on 
given heuristics, \eg, a large amount of profits are made in a very short time. We 
further applied \tool in several popular \defi projects on \ethereum and managed to 
uncover potential attacks which were previously unreported.

	\section{Attack Detection For DeFi}
	\label{sec:method}
	\subsection{Overview}
\label{subsec:overview}

The general workflow of \tool is shown in Figure \ref{fig:framework}. 
Specifically, \tool works in a two-phase manner. In the first phase, 
\tool performs symbolic analysis on smart contracts of a given \defi 
project. This is realized by extending an underlying smart contract 
analyzer \textsc{Seraph}~\cite{yang2020seraph}, which is also developed by our team. 
Specifically, the goal of this phase is to model the inter-DeFi oracle dependency, 
\ie, how does the oracle data provided by one \defi affect services of another. 
In cases where oracle-dependent state updates are found, we identify the \defi 
as potentially vulnerable. Next, \tool installs a runtime 
monitor in the second phase for vulnerable \defi projects to detect external attacks. 
Specifically, \tool uses a transaction monitor to collect related 
transactions based on extracted features, \eg, address. Then, 
end-to-end transactions are analyzed according to 
predefined heuristics, \eg, a large profit is made in a short period. 
Potential attacks are flagged by \tool when an abnormal sequence of 
transactions is detected. Moreover, \tool generates analysis report to 
help blockchain service providers diagnose the found problems.

\subsection{Oracle Analysis}
\label{subsec:connector}

As aforementioned, \tool performs \emph{oracle analysis} to check whether 
a \defi is dependent on the oracle provided by another \defi. Particularly, 
we focused on the price feed of assets shared through oracles.

\begin{figure}[h]
\begin{lstlisting}[frame=tblr, language=Solidity, showstringspaces=false]
function calculateContinuousMintReturn(uint _amount) 
  public view returns (uint mintAmount) {
	return CURVE.calculatePurchaseReturn(totalSupply(), 
	reserveBalance, uint32(reserveRatio), _amount);
}

function sell(uint _amount, uint _min) external 
  returns (uint _bought) {
  _bought = _sell(_amount);
  require(_bought >= _min, "slippage");
  _burn(msg.sender, _amount);
  DAI.transfer(msg.sender, _bought);
  ...
}
\end{lstlisting}
\caption{Oracle in the \texttt{EMN} project}
\label{lst:oracle}
\end{figure}

An illustrative example of \texttt{EMN} project is given in Figure~\ref{lst:oracle}. Specifically, 
the function call at line 9 implicitly invokes the function from line 1--5, 
which receives an oracle at line 3--4. Moreover, the payment at line 
12 is dependent on the oracle due to a data flow from line 9 to 12. 
That said, \texttt{EMN} has an oracle-dependent state update in its smart contracts. 
To enable such oracle analysis, \tool extended the \textsc{Seraph} smart contract 
analyzer to perform symbolic reasoning on oracles. Specifically, when processing a 
\texttt{CALL} instruction to a specified oracle, \tool starts a data flow analysis to 
track whether the value retrieved from the oracle is linked to a further state operation, 
\eg, payment, storage update \etc. In cases where a feasible link is detected as in 
Figure~\ref{lst:oracle}, \tool dumps the data flow and identifies given \defi 
as potentially vulnerable.

\subsection{Automatic Attack Monitoring}
\label{subsec:ssg}

For vulnerable \defi projects, \tool launches a runtime transaction 
monitoring to detect external attacks. To this end, \tool allows users 
to specify targeted projects and further performs end-to-end analysis 
on relevant transactions. In general, the analysis aims at finding 
violations on invariant as predefined heuristic rules.

Specifically, as in Figure~\ref{fig:bzx-attack} with associated 
\defi projects \texttt{Uniswap}, \texttt{Synthetix} and \texttt{Kyber}, 
\tool first marks a random transaction $t_0$ in these platforms as a target. 
Moreover, we search for other related transactions $t_1 \cdots t_k$ based 
on the sender address $x$ in $t_0$. Additionally, we filter transactions which 
are not in the same block as $t_0$ in order to find frequent transactions, 
which are more likely to be involved in attacks. With the collection of 
$t_0\cdots t_k$ transactions, \tool runs a process to calculate the benefits 
received by $x$ and its cost as well. With both numbers, \tool is 
then able to determine whether $x$ is attacking the target \defi by comparing 
the profit made by $x$ and a threshold value as configured. \tool will also 
dumps the malicious sequence of transactions to facilitate an in-depth analysis 
of the potential attack.

	\section{Design of \tool}
	\label{sec:design}
	\subsection{Architecture}
\label{subsec:architecture}

\begin{figure}[h]
	\centering
	\includegraphics[width=\linewidth]{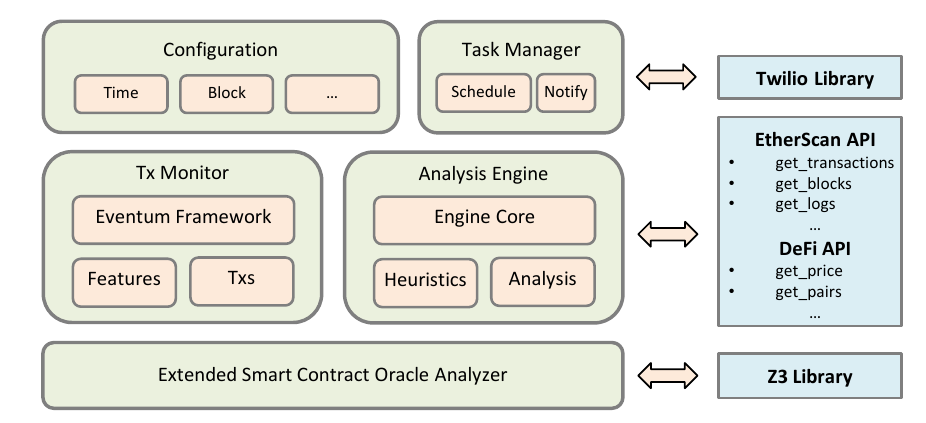}
	\caption{\label{fig:architecture}%
		The general architecture of \tool.}
\end{figure}

The \tool is implemented as a web platform with front and back-end services, where the back-end architecture is shown in Figure \ref{fig:architecture}.
There are five functional modules in this architecture. At the bottom, 
\tool extends a smart contract analyzer to perform oracle analysis as introduced 
earlier. \texttt{Z3}~\cite{z3} is adopted as the SMT solver in this module. 

In the middle are \emph{Tx Monitor} and \emph{Analysis Engine}. Transactions 
are monitored and collected via the \texttt{Eventum} framework, which streams events 
from blockchain to \tool. Moreover, we implemented the analysis engine 
to detect potential attacks based on collected transactions and events. 
At the top layer, \tool provides a \emph{Configuration} module to allow users to specify
detection criteria, \eg, in physical time or block number. Furthermore, the 
\emph{Task Manager} module is designed to schedule detection tasks submitted from front-end and send back notifications to users with the \texttt{Twillo} library.

\subsection{Main Functionalities}
\label{subsec:funcitonality}

\begin{figure}[htbp]
	\centering
	\includegraphics[width=0.9\linewidth]{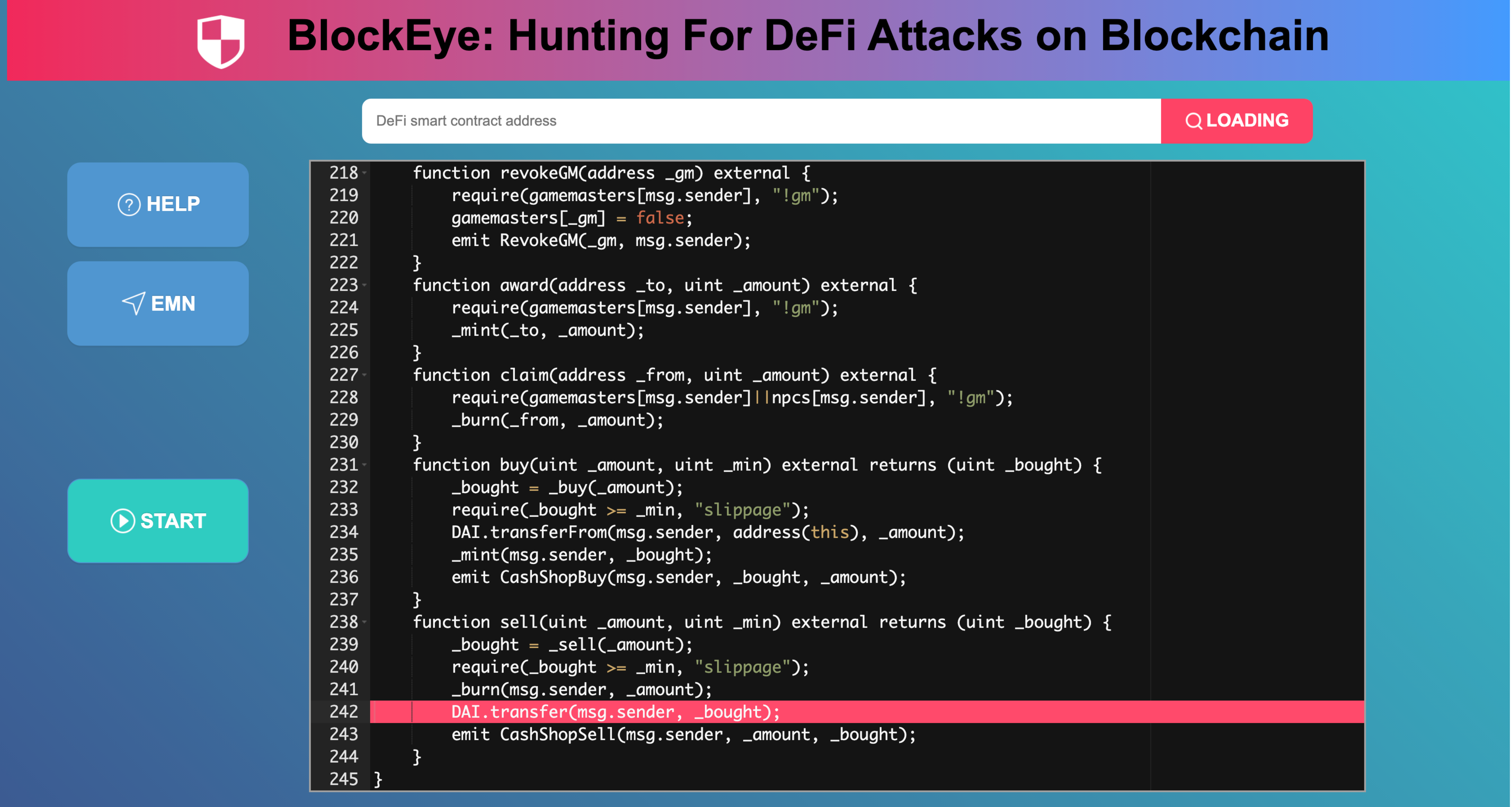}
	\caption{\label{fig:screenshot-input}%
	The input interface for \tool.}	
\end{figure}
\begin{figure}[htbp]
	\centering
	\includegraphics[width=0.9\linewidth]{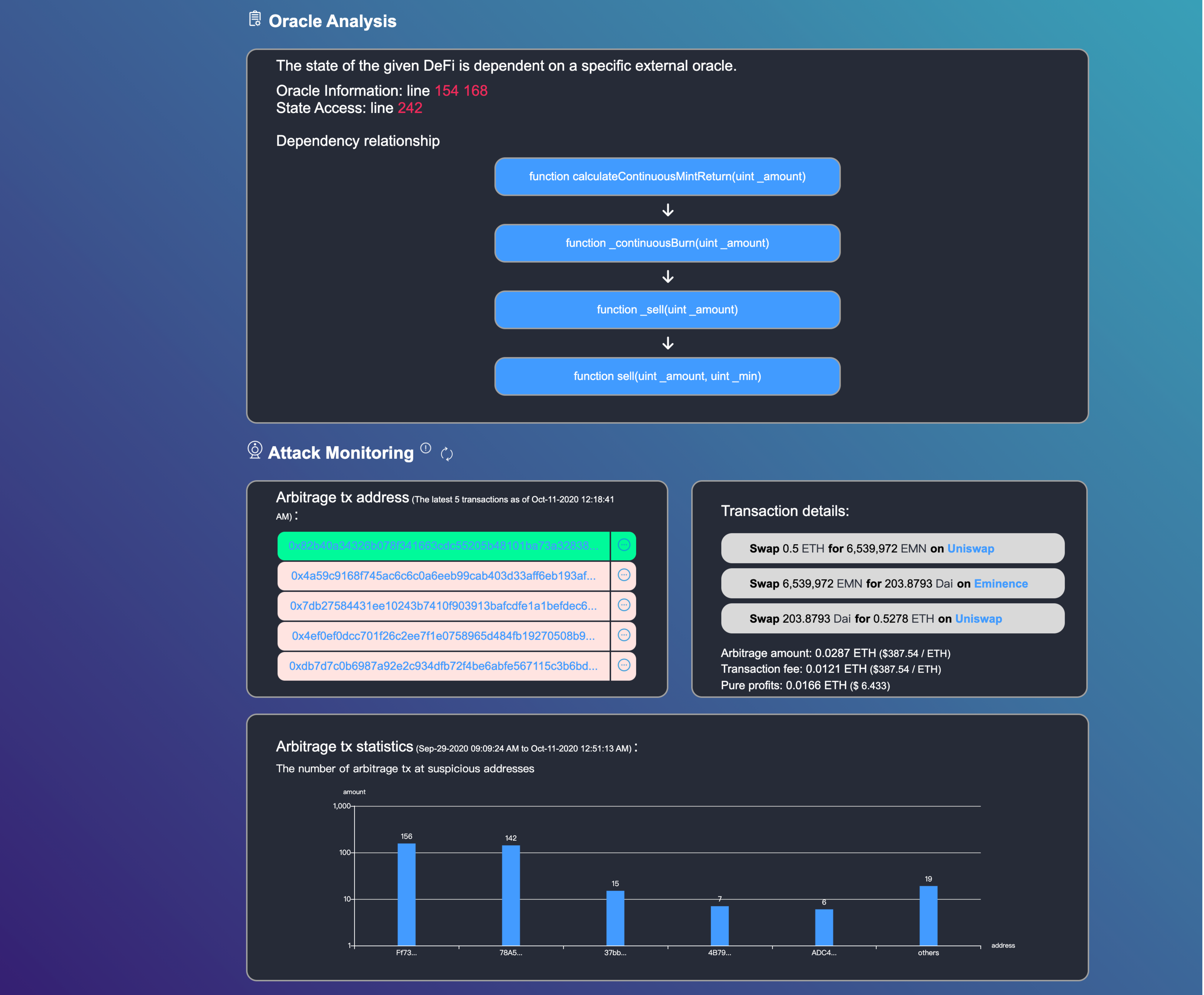}
	\caption{\label{fig:screenshot-output}%
	The output interface for \tool.}
\end{figure}

We now describe the input and output interfaces of \tool with screenshots shown in Figure \ref{fig:screenshot-input} and Figure \ref{fig:screenshot-output}.

As in Figure \ref{fig:screenshot-input}, \tool expects \defi smart contract source code as input.
Users can either type in code in the code editor, or provide the address of a deployed \defi project.
\tool then will try to load corresponding source code using Etherscan's source code retrieving API.
Once smart contract code is available, users are free to click the \emph{START} button to launch security analysis on the given \defi project.

An example output of \tool is in Figure \ref{fig:screenshot-output}.
Here, results are divided into two parts: \emph{Oracle Analysis}, which shows potential oracle dependencies found in \defi source code, and \emph{Attack Monitoring}, that provides information of real-world attack transactions which break heuristic invariants as described in Section \ref{subsec:ssg}.
For example, in Figure \ref{fig:screenshot-output}, \tool has found an oracle dependency which spans across four smart contract functions, with oracle contract defined in code line $154$ and fired by function \texttt{calculateContinuousBurnReturn} in line $168$.
Corresponding state access operation is in line $242$ as a request to transfer \texttt{DAI} with dependent amount of value.
Besides, in Figure \ref{fig:screenshot-output}, \tool shows a list of five latest suspicious transactions, each with detailed information on its internal operation.
At last, \tool also presents a graph of top attackers along with their number of detected attack transactions, which may help users in further investigations.


	
	\section{Preliminary Evaluation}
	\label{sec:eval}
	We conducted a preliminary evaluation on \tool to validate its effectiveness in finding oracle-dependent state updates.
Specifically, we considered eight DeFi projects on Ethereum, i.e., bZx, DDEX, Aave, dYdX, Compound, Nuo, Oasis, and Eminence.

In Table \ref{tab:tool_comparison}, we present a comparison of \tool with Codefi Inspect~\cite{codefi} 
in oracle-dependent state update detection.
The results show that \tool successively identifies all vulnerable DeFis with no false positive or false negative alert.
Whereas, Codefi Inspect falsely ignores vulnerabilities in DDEX, leading to a false negative (FN) result.

\begin{table}[htbp]
	\centering
	\caption{A comparison of \tool and Codefi Inspect in oracle-dependent state update detection. TP: True Positive; TN: True Negative; FN: False Negative; N/A: Not Available.}\label{tab:tool_comparison}
	
	\resizebox{0.66\columnwidth}{!}{%
	\begin{threeparttable}
		\begin{tabularx}{0.73\linewidth}{c c c}
			\toprule
			\multicolumn{1}{c}{\bf DeFi} & \multicolumn{1}{c}{\bf Codefi Inspect} & \multicolumn{1}{c}{\bf \tool}\\
			
			\midrule
			bZx & TP & TP\\
			DDEX & FN & TP\\
			Aave & TN & TN\\
			dYdX & TN & TN\\
			Compound & TN & TN\\
			Nuo & N/A & TN\\
			Oasis & N/A & TN\\
			Eminence & N/A & TP\\
			
			\bottomrule
		\end{tabularx}
	\end{threeparttable}%
	}
\end{table}

We further evaluated \tool with real-world transactions on the Ethereum mainnet.
In Table \ref{tab:detailed_results}, we show detailed results of detected arbitrage transactions in two DeFis, i.e., ETH/sUSD token pair on bZx and DAI/EMN on Eminence.

\begin{table}[htbp]
	\centering
	\caption{Detailed results of \tool with two DeFis.}\label{tab:detailed_results}
	
	\resizebox{0.95\columnwidth}{!}{%
	\begin{threeparttable}
		\begin{tabularx}{1.05\linewidth}{c r l l r l l}
			\toprule
			\multicolumn{1}{c}{\bf DeFi} & \multicolumn{3}{c}{\bf bZx(ETH/sUSD)} & \multicolumn{3}{c}{\bf Eminence(DAI/EMN)}\\
			
			\midrule
			Block & \multicolumn{3}{c}{10799704 $\sim$ 10950575} & \multicolumn{3}{c}{10956504 $\sim$ 11031087}\\
			\# Tx & \multicolumn{3}{c}{$\approx$ 7138} & \multicolumn{3}{c}{$\approx$ 1486}\\
			\multirow{3}{*}{\# Suspicious Tx} & slippage & > 0.05 & 25 & slippage & > 0.05 & 124\\
			& slippage & > 0.07 & 23 & slippage & > 0.057 & 107\\
			& slippage & > 0.1 & 19 & slippage & > 0.059 & 37\\
			
			\bottomrule	
		\end{tabularx}

	\end{threeparttable}%
	}
	
\end{table}

For example, as for bZx (ETH/sUSD), there were around $7,138$ valid transactions 
detected between block $10,799,704$ and $10,950,575$.
By enforcing different slippage thresholds, \tool found $19$ to $25$ suspicious arbitrage transactions, 
\eg, $19$ for slippage threshold $0.1$ and $25$ for slippage threshold $0.05$.
Besides, for Eminence(DAI/EMN), the number of suspicious transactions found ranged 
from $37$ to $124$ with different slippage thresholds, where overall valid transactions detected 
were $1,486$ between block $10,956,504$ and $11,031,087$.

	\section{Related Work}
	\label{sec:rw}
	
Security problems of smart contracts have been widely discussed 
in recent years~\cite{liu2018reguard,liu2018s,tsankov2018securify,Luu2016Making,yang2020seraph}. 
Luu~\etal highlighted four types of vulnerabilities 
for smart contracts~\cite{Luu2016Making}. 
Tsankov~\etal proposed a verification technique~\cite{tsankov2018securify}, which transforms 
\ethereum smart contracts into Datalog logics~\cite{eiter1997disjunctive}. 
Permenev~\etal further presented their solution to verify smart contracts in an inductive 
manner~\cite{permenev2019verx}. In addition to security problems, 
Liu~\etal proposed a statistical approach to identify potential code 
smells~\cite{liu2018s}. Security of \defi projects is relatively less 
discussed in previous works. Several mathematical and economic models 
were proposed to help understand risks of \defi in a theoretical 
manner~\cite{qin2020attacking,kamps2018moon,liu2020first,gudgeon2020decentralized}. 
	
	\section{Conclusion}
	\label{sec:conclusion}
	In this paper, we highlighted \tool as an open platform 
to detect \defi attacks on blockchain. Compared to existing 
analyzers for smart contracts, \tool provides important 
capabilities to model dependency among \defi projects and 
flag potential end-to-end attacks at real-time. The key insights 
behind \tool are symbolic oracle analysis and pattern-based 
runtime transaction validation. We applied \tool in several 
popular \defi projects on \ethereum and managed to find 
potential attacks previously unreported.


    \section{Data Availability Statement}
	\label{sec:Availability}
	For ethical considerations, experimental data used in our work will be publicly available after discussions with the relevant DeFi development team.


	\bibliographystyle{IEEEtran}
	\bibliography{blockeye}

\begin{thebibliography}{10}
\providecommand{\url}[1]{#1}
\csname url@samestyle\endcsname
\providecommand{\newblock}{\relax}
\providecommand{\bibinfo}[2]{#2}
\providecommand{\BIBentrySTDinterwordspacing}{\spaceskip=0pt\relax}
\providecommand{\BIBentryALTinterwordstretchfactor}{4}
\providecommand{\BIBentryALTinterwordspacing}{\spaceskip=\fontdimen2\font plus
\BIBentryALTinterwordstretchfactor\fontdimen3\font minus
  \fontdimen4\font\relax}
\providecommand{\BIBforeignlanguage}[2]{{%
\expandafter\ifx\csname l@#1\endcsname\relax
\typeout{** WARNING: IEEEtran.bst: No hyphenation pattern has been}%
\typeout{** loaded for the language `#1'. Using the pattern for}%
\typeout{** the default language instead.}%
\else
\language=\csname l@#1\endcsname
\fi
#2}}
\providecommand{\BIBdecl}{\relax}
\BIBdecl

\bibitem{Wood2014Ethereum}
G.~Wood, ``Ethereum: A secure decentralised generalised transaction ledger,''
  \emph{Ethereum Project Yellow Paper}, vol. 151, 2014.

\bibitem{Luu2016Making}
L.~Luu, D.-H. Chu, H.~Olickel, P.~Saxena, and A.~Hobor, ``Making smart
  contracts smarter,'' in \emph{Proceedings of the 2016 ACM SIGSAC Conference
  on Computer and Communications Security}.\hskip 1em plus 0.5em minus
  0.4em\relax ACM, 2016, pp. 254--269.

\bibitem{tsankov2018securify}
P.~Tsankov, A.~Dan, D.~D. Cohen, A.~Gervais, F.~Buenzli, and M.~Vechev,
  ``Securify: Practical security analysis of smart contracts,'' \emph{arXiv
  preprint arXiv:1806.01143}, 2018.

\bibitem{liu2018reguard}
C.~Liu, H.~Liu, Z.~Cao, Z.~Chen, B.~Chen, and B.~Roscoe, ``Reguard: finding
  reentrancy bugs in smart contracts,'' in \emph{ICSE (Companion)}.\hskip 1em
  plus 0.5em minus 0.4em\relax ACM, 2018, pp. 65--68.

\bibitem{liu2018s}
H.~Liu, C.~Liu, W.~Zhao, Y.~Jiang, and J.~Sun, ``S-gram: towards semantic-aware
  security auditing for ethereum smart contracts,'' in \emph{ASE}.\hskip 1em
  plus 0.5em minus 0.4em\relax ACM, 2018, pp. 814--819.

\bibitem{yang2020seraph}
Z.~Yang, H.~Liu, Y.~Li, H.~Zheng, L.~Wang, and B.~Chen, ``Seraph: enabling
  cross-platform security analysis for evm and wasm smart contracts,'' in
  \emph{Proceedings of the ACM/IEEE 42nd International Conference on Software
  Engineering: Companion Proceedings}, 2020, pp. 21--24.

\bibitem{z3}
``Microsoft z3 smt solver,'' \url{https://z3.codeplex.com/}, 2019.

\bibitem{codefi}
``Codefi inspect,'' \url{https://inspect.codefi.network/}, 2020.

\bibitem{eiter1997disjunctive}
T.~Eiter, G.~Gottlob, and H.~Mannila, ``Disjunctive datalog,'' \emph{ACM
  Transactions on Database Systems (TODS)}, vol.~22, no.~3, pp. 364--418, 1997.

\bibitem{permenev2019verx}
A.~Permenev, D.~Dimitrov, P.~Tsankov, D.~Drachsler-Cohen, and M.~Vechev,
  ``Verx: Safety verification of smart contracts,'' \emph{Security and
  Privacy}, vol. 2020, 2019.

\bibitem{qin2020attacking}
K.~Qin, L.~Zhou, B.~Livshits, and A.~Gervais, ``Attacking the defi ecosystem
  with flash loans for fun and profit,'' \emph{arXiv preprint
  arXiv:2003.03810}, 2020.

\bibitem{kamps2018moon}
J.~Kamps and B.~Kleinberg, ``To the moon: defining and detecting cryptocurrency
  pump-and-dumps,'' \emph{Crime Science}, vol.~7, no.~1, p.~18, 2018.

\bibitem{liu2020first}
B.~Liu and P.~Szalachowski, ``A first look into defi oracles,'' \emph{arXiv
  preprint arXiv:2005.04377}, 2020.

\bibitem{gudgeon2020decentralized}
L.~Gudgeon, D.~Perez, D.~Harz, A.~Gervais, and B.~Livshits, ``The decentralized
  financial crisis: Attacking defi,'' \emph{arXiv preprint arXiv:2002.08099},
  2020.

\end{thebibliography}

\end{document}